\documentclass[runningheads]{chaos2009}

\usepackage{chaos2009References} % For References style

\usepackage{graphicx} % standard LaTeX graphics tool
\begin{document}
%
% \title* lets you specify the title of your manuscript.
% Use \protect\newline to force a line break in your title.
\title*{Introducing Chaos in \protect\newline Economic Gas-Like Models}
%
% \toctitle specifies the title as will be printed in the table of
% contents.
% Use \protect\newline to force a line break in your title.
\toctitle{Introducing Chaos in Economic Gas-Like Models}
%
% \titlerunning defines the title in the running head. Abbreviate
% your title, if the full title is too long to fit in the running
% head.
\titlerunning{Chaos in Economic Gas-Like Models}
%
% \authors specifies the authors. Please use initials. Authors are
% seperated by the \and command. Use the \inst{1} and \inst{2} commands
% to define the reference mark to your affiliation if needed.
\author{
  Carmen Pellicer-Lostao and Ricardo L\'{o}pez-Ruiz%\inst{1}
%  \and
%  Ricardo L\'{o}pez-Ruiz\inst{2}
}
%
% The following command allow each of the authors to appear
% in the author index.
% \index{Author, A.}
\index{Pellicer-Lostao, C.}
\index{L\'{o}pez-Ruiz, R.}

%
% \authorrunning specifies the author name(s) in the running head.
% If there are more than two authors, please abbreviate author list
% (e.g., Skiadas  et al.) for running head.
\authorrunning{C. Pellicer-Lostao and R. L\'{o}pez-Ruiz}
%
% The \institute command lets you specify  your affiliation and
% your address. Seperate two or more different affiliations by the
% \and command.
\institute{
  Department of Computer Science and\\
  Institute for Biocomputation and Physics of Complex Systems (BIFI)\\
  University of Zaragoza\\
  5004 Zaragoza, Spain\\
  (e-mail: {\tt carmen.pellicer@unizar.es} and {\tt rilopez@unizar.es})
%   \and
%  (e-mail: {\tt rilopez@unizar.es})
}

% Typeset the title
\maketitle

\begin{abstract}
This paper considers ideal gas-like models of trading markets, where each agent is identified
 as a gas molecule that interacts with others trading in elastic or money-conservative collisions.
  Traditionally, these models introduce different rules of random selection and exchange between pair agents.
   Unlike these traditional models, this work introduces a chaotic procedure able of breaking the pairing
   symmetry of agents $(i,j)\Leftrightarrow(j,i)$. Its results show that, the asymptotic money distributions 
   of a market under chaotic evolution can exhibit a transition from Gibbs to Pareto distributions, 
   as the pairing symmetry is progressively broken. \keyword{Complex Systems} 
   \keyword{Chaos} \keyword{Econophysics} \keyword{Gas-like Models}
\keyword{Money Dynamics} \keyword{Chaotic Simulation}
\end{abstract}

\section{Introduction}

Modern Econophysics is a relatively new discipline ~\cite{Mantegna} that applies many-body techniques developed
in statistical mechanics to the understanding of self-organizing economic systems ~\cite{yakovenko2007}.  The
techniques used in this field ~\cite{yakovenko2000},~\cite{chacrabarti2000}, ~\cite{bouchaud} have to do with
agent-based models and simulations. The statistical distributions of money, wealth and income are obtained on a
community of agents under some rules of trade and after an asymptotically high number of interactions between
the agents.

The conjecture of a kinetic theory of (ideal) gas-like model for trading markets was first discussed 
in 1995 ~\cite{chacrabarti1995} by econophysicists. This model consi-ders a closed economic community 
of individuals where each agent is identified as a gas molecule that interacts randomly with others, 
trading in elastic or money-conservative collisions. The interest of this model is that, by analog with 
energy, the equilibrium probability distribution of money follows the exponential Boltzmann-Gibbs law for 
a wide variety of trading rules ~\cite{yakovenko2007}.

This result is coherent with real economic data in some capitalist countries up to some extent, for 
in high ranges of wealth evidences are shown of heavy-tail distributions ~\cite{yakovenko2001}, ~\cite{forbes2006}. 
Different reasons can be argued for this failure of the gas-like model. In this work, the authors suppose 
that real economy is not purely random.

On one hand, there is some evidence of markets being not purely random. Real economic transactions are 
driven by some specific interest (or profit) between the different interacting parts. On the other hand, 
history shows the unpredictable component of real economy with its recurrent crisis. Hence, it can be 
sustained that the short-time dynamics of economic systems evolves under deterministic forces and, 
in the long term, these systems display inherent unpredictability and instability. Therefore, 
the prediction of the future situation of an economic system resembles somehow to the weather prediction. 
It can be concluded that determinism and unpredictability, the two essential components of chaotic systems, 
take part in the evolution of Economy and Financial Markets.

Consequently, one may consider of interest to introduce chaotic patterns in the theory of (ideal) 
gas-like model for trading markets. One can observe this way, which money distributions are obtained, 
how they differ from the referenced exponential distribution and how they resemble real economic distributions.

The paper presented here, follows precisely this approach. It focuses on the statistical distribution of money
in a closed community of individuals, where agents exchange their money under a certain conservative rule. But
unlike these traditional models, this work is going to introduce chaotic trade interactions. More specifically
it introduces a chaotic procedure for the selection of agents that interact at each transaction. This chaotic
selections of trading partners is going to determine the success of some individuals over others. In the end
it will be seen that, as in real life, a minority of chaos-predilected people can follow heavy tail
distributions.

The contents of this paper are organized as follows: section 2 describes the simulation scenario. 
Section 3 shows the results obtained in this scenario. Final conclusions are discussed in Section 4.

\section{Scenario of Chaotic Simulation}

The simulation scenario considered here follows a traditional gas-like model, but the rules of trade 
intend to be less random and more chaotic. The study of these scenarios was first proposed by the authors 
in \cite{LNCS}. There, it is shown that the use of chaotic numbers produces the exponential as well as 
other wealth distributions depending on how they are injected to the system. This paper considers the 
scenario where the selection of agents is chaotic, while the money exchanged at each interaction is a 
random quantity.

In the computer simulations presented here, a community of $N$ agents is given with an initial equal 
quantity of money, $m_0$, for each agent. The total amount of money, $M=N*m_0$, is conserved. For each 
transaction, at a given instant $t$, a pair of agents $(i,j)$ is selected chaotically and a random amount 
of money $\Delta m$ is traded between them.

To produce chaotically a pair of agents $(i,j)$ for each interaction, a 2D chaotic system is considered. 
The pair $(i,j)$ is easily obtained from the coordinates of a chaotic point at instant $t$, $X_t=[x_t, y_t]$, 
by a simple float to integer conversion ($x_t$ and $y_t$ to $i$ and $j$, respectively). Additionally, 
a random number from a standard random generator is used to obtain a float number $\upsilon$ in the 
interval $[0,1]$. This number produces the random the quantity of money $\Delta m$ traded between 
agents $x_t$ and $y_t$.

The particular rule of trade is the following: let us consider two agents $i$ and $j$ with their respective
wealth, $m_i$ and $m_j$ at instant $t$. At each interaction, the quantity $\Delta m=\upsilon*(m_i+m_j)/2$, 
is taken from $i$ and given to $j$ . Here, the transaction of money is quite asymmetric as agent $j$ is 
the absolute winner, while $i$ becomes the looser. If $i$ has not enough money, no transfer takes place. 
This rule is selected
for it has been extensively used and so, comparations can be established with popularly referenced 
literature ~\cite{yakovenko2007}.

The particular 2D chaotic system used in the simulations is the model (a) in ~\cite{ric}. 
This system is obtained by a multiplicative coupling of two logistic maps. 
A real-time animation of this system can be seen in ~\cite{mathematica}, where $(x_t,y_t)= T(x_{t-1},y_{t-1})$. 
This system is given by the following equation: $T:[0,1]\times[0,1]\longrightarrow[0,1]\times[0,1]$

\begin{equation}
\label{system}
x_t =\lambda_a(3y_{t-1}+1)x_{t-1}(1-x_{t-1}),\hskip 8mm  y_t=\lambda_b(3x_{t-1}+1)y_{t-1}(1-y_{t-1}).
\end{equation}

For each transaction at a given instant t, two chaotic floats in the interval $[0,1]$ are produced. 
These values are used to obtain $i$ and $j$ through the following equation:

\begin{equation}
\label{chaoAgents}
i=(int)(x_t*N),\hskip 8mm  j=(int)(y_t*N).
\end{equation}

\begin{figure}
  \centerline{
    \includegraphics[width=0.5\textwidth]{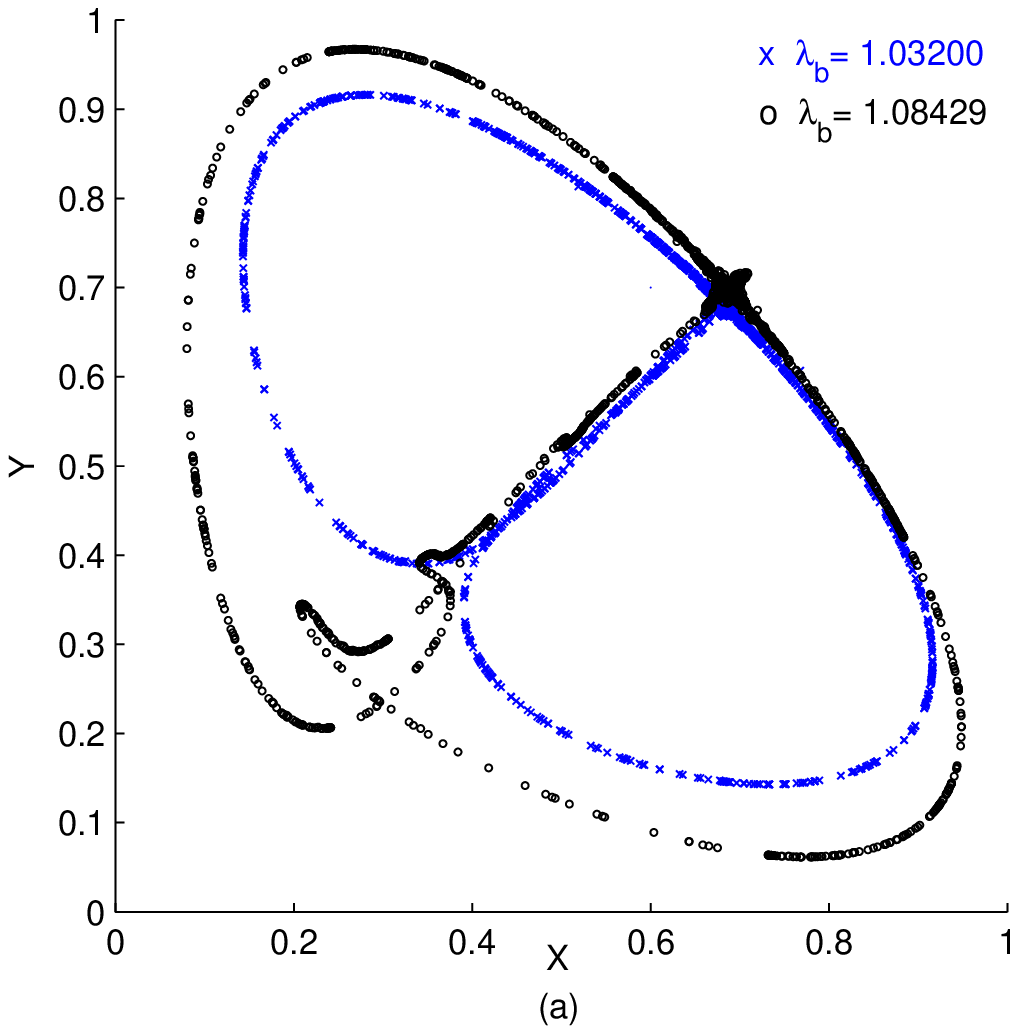} \hskip 3mm \includegraphics[width=0.5\textwidth]{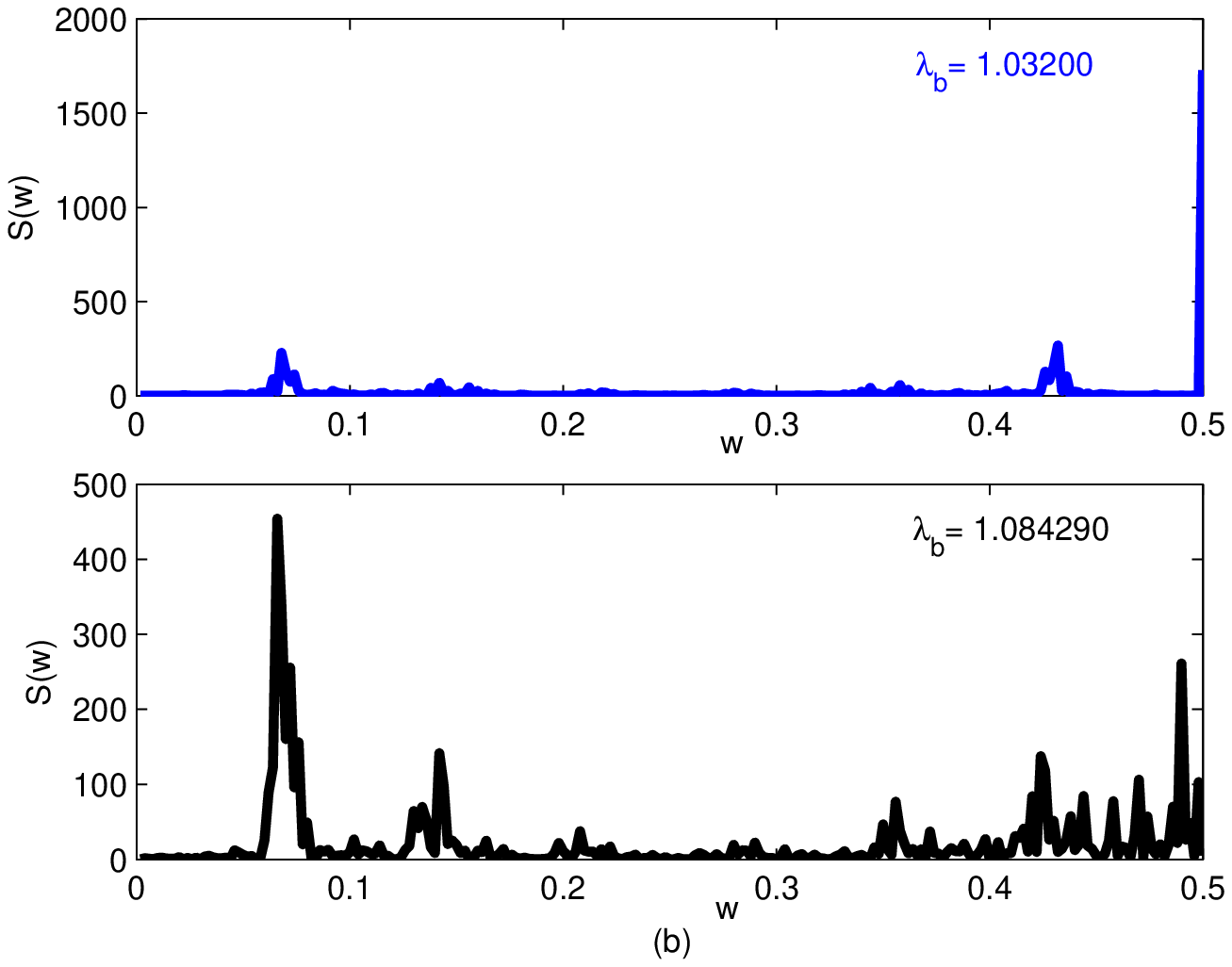}}
  \caption{In (a) representation of 2000 points of the chaotic attractor of the Logistic Bimap.
  In (b) representation of the spectrum of coordinate $x_t$. The values of the parameters 
  are $\lambda_a= \lambda_b= 1.032$ in the black curve and $\lambda_a=1.032$, $\lambda_b=1.084290$ 
  in the blue curve.}
  \label{fig1}
\end{figure}

From a geometrical point of view, this Logistic Bimap presents a chaotic attractor in the interval 
$\lambda_{a,b}\in[1.032, 1.0843 ]$. The selection of this system is due to the fact that its symmetry 
can be adjusted as desired through the proper selection of parameters $\lambda_a$ and $\lambda_b$. 
This can be observed in Fig. \ref{fig1}.

In fact, the system is symmetric respective to the diagonal when $\lambda_a= \lambda_b$. 
The spectrum of coordinate $x_t$ also shows a peak for $w=0.5$, presenting an oscillation of 
period two that makes it jump over the diagonal axis alternatively between consecutive points in time. 
Both sub-spaces $x>y$ and $y>x$ are visited with the same frequency and the shape of the attractor is symmetric.

When $\lambda_b$ becomes greater than $\lambda_a$ the part of the attractor in sub-space $x<y$ becomes wider
 and the frequency of visits of each sub-space becomes different. This is going to be particularly interesting 
 for our purposes, as the degree of symmetry of the chaotic system is going to be an input variable in the 
 simulations.

\section{Chaotic Wealth Distributions}

A community of $N=5000$ agents with initial money of $m_0=\$1000$ is taken. The Logistic Bimap variables $x_t$
and $y_t$ in chaotic regime will be used as simulation parameters to obtain trading agents $i$ and $j$. 
The simulations take a total time of $T= 2*N^2= 50$ millions of transactions.

Different cases are considered, as different values of the chaotic parameters $\lambda_a$ and $\lambda_b$ are
used. In this way the symmetry of the selection of agents is going to vary from total symmetry to the highest
value of asymmetry, as it was shown in Fig. \ref{fig1}. Table \ref{table1} shows the values of parameters used 
for the different simulations:

\begin{table} [h]
\label{table1}
\begin{center}
\begin{tabular}{|c|c|c|c|c|c|c|c|c|}
   \hline
  CASE & $1$ & $2$ & $3$ & $4$ & $5$ & $6$ & $7$ & $8$\\
  \hline
  $\lambda_a$ & $1.032$ & $1.032$ & $1.032$ & $1.032$ & $1.032$ & $1.032$ & $1.032$ & $1.032$\\
  \hline
  $\lambda_b$ & $1.032$ & $1.03781$ & $1.04362$ & $1.049430$ & $1.06105$ & $1.07267$ & $1.07848$ & $1.08429$\\
  \hline
\end{tabular}
\end{center}
\caption{List of values of $\lambda$ parameter used in the Logistic Bimap for the chaotic simulations.}
\end{table}

\begin{figure}
  \centerline{
    \includegraphics[width=0.5\textwidth]{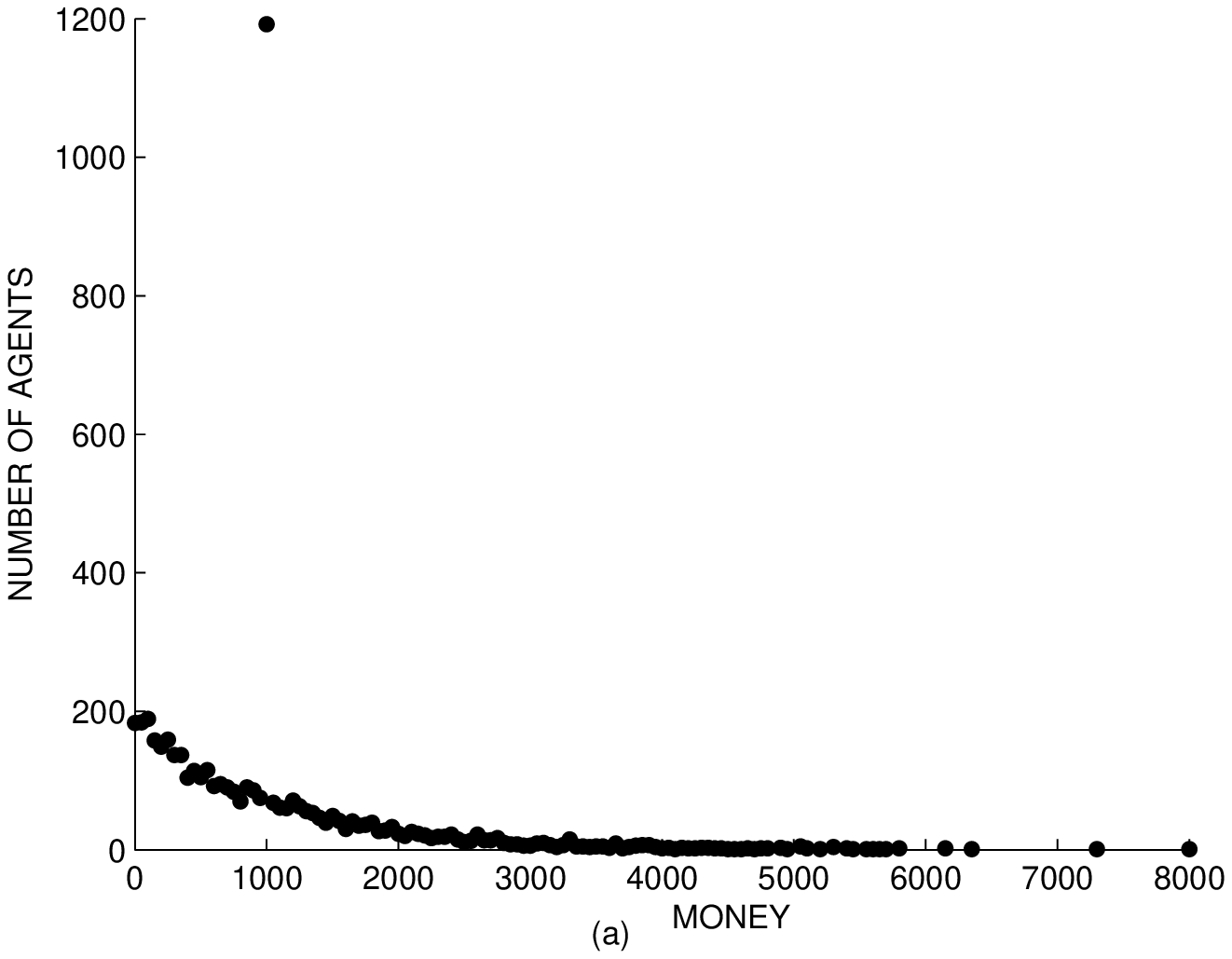} \hskip 3mm 
	\includegraphics[width=0.5\textwidth]{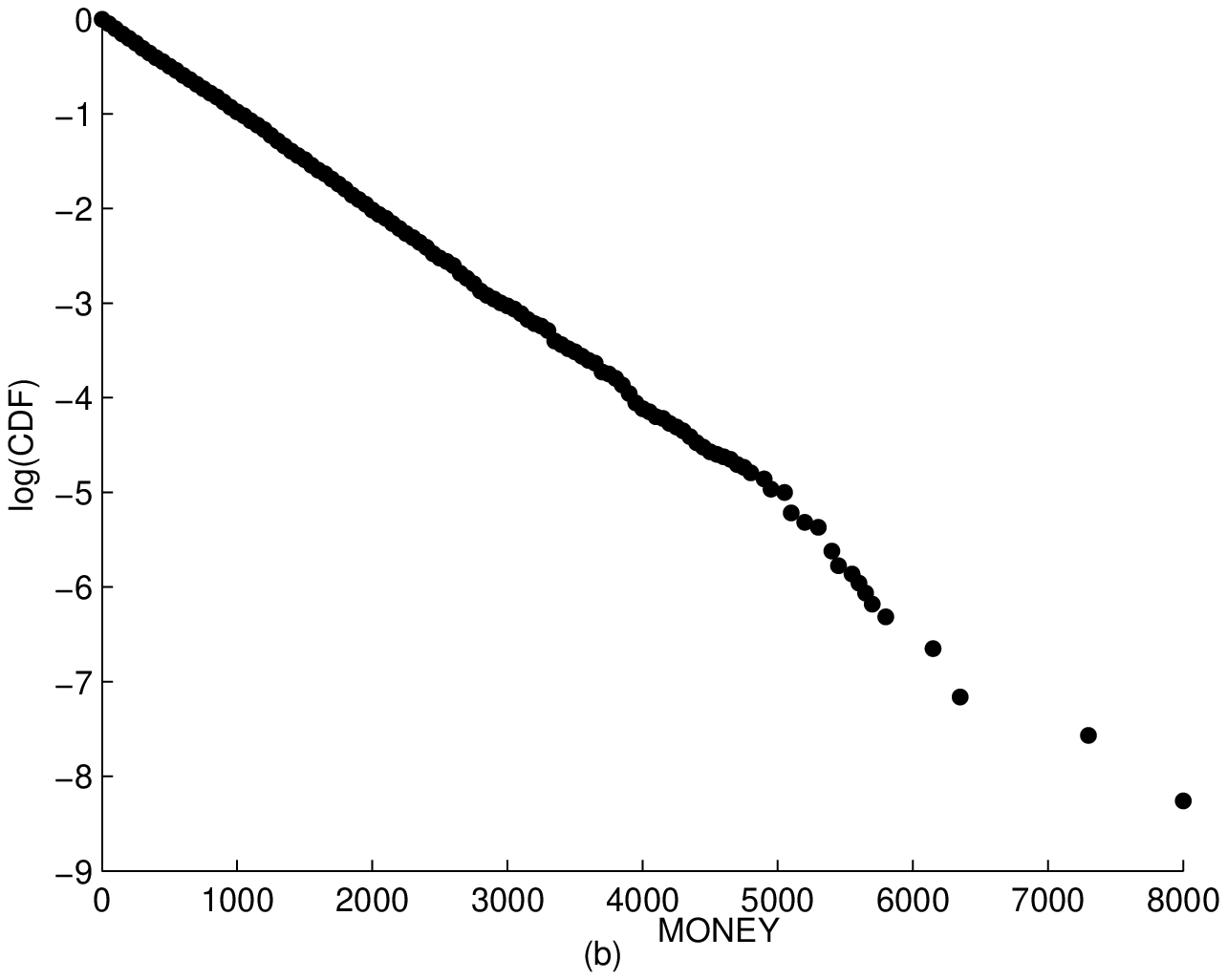}}
  \caption{Representations of the final money distribution obtained for chaotic parameters 
  $\lambda_a= \lambda_b=1.032$. In (a) the distribution with $5000$ agents. A peak can be seen at 
  $\$1000$ with the passive agents. In (b) the cumulative distribution function (CDF) is drawn for the 
  community of real participants ($3867$ agents).}
  \label{fig2}
\end{figure}

The resulting money distributions are then obtained as the $\lambda_b$ varies from $1.032$ to $1.08429$. 
In Fig. \ref{fig2}(a), the wealth distribution for the symmetric case is presented. As it can be seen it 
resembles an exponential distribution. Another interesting point appears in this case. This is the high number 
of individuals ($1133$ agents) that keep their initial money in Fig. \ref{fig2}(a). The reason is that they 
don't exchange money at all. The chaotic numbers used to choose the interacting agents are forcing trades 
between a deterministic group of them, and hence some trading relations result restricted.

When the passive agents are removed of the model, one can obtain the money distribution of the interacting 
agents. Fig. \ref{fig2}(b) shows the cumulative distribution function (CDF) obtained for the symmetric case. 
Here the proba-bility of having a quantity of money bigger or equal to the variable MONEY, is depicted in natural 
log plot, showing clearly the exponential distribution.

When $\lambda_b$ varies from $1.032$ to $1.08429$ it is observed that the number of non-participants decreases. 
This is because the chaotic map expands (see Fig. \ref{fig1}) and its resulting projections on axis $x$ and $y$ 
grow in range, taking a greater group of $i$ and $j$ values when equation (\ref{chaoAgents}) is computed. 
Taking these non-participants off the final money distributions, and so their money too, one can obtain the 
final CDF's for the different values of $\lambda_b$.  When these distributions are depicted an interesting 
progression is shown. As $\lambda_b$ increases, these distributions diverge from the exponential shape.

\begin{figure} [h]
  \centerline{
    \includegraphics[width=0.5\textwidth]{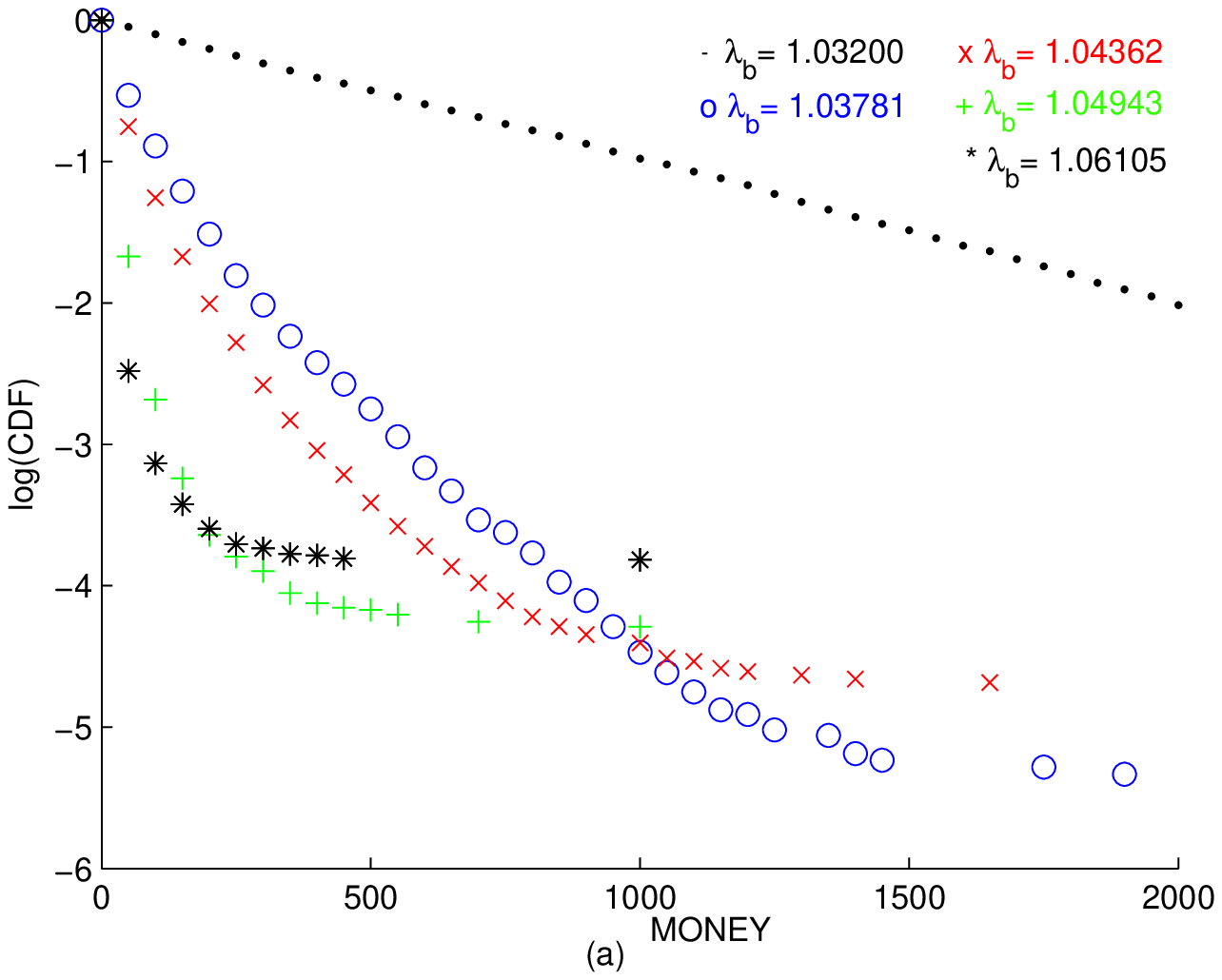} \hskip 3mm 
	\includegraphics[width=0.5\textwidth]{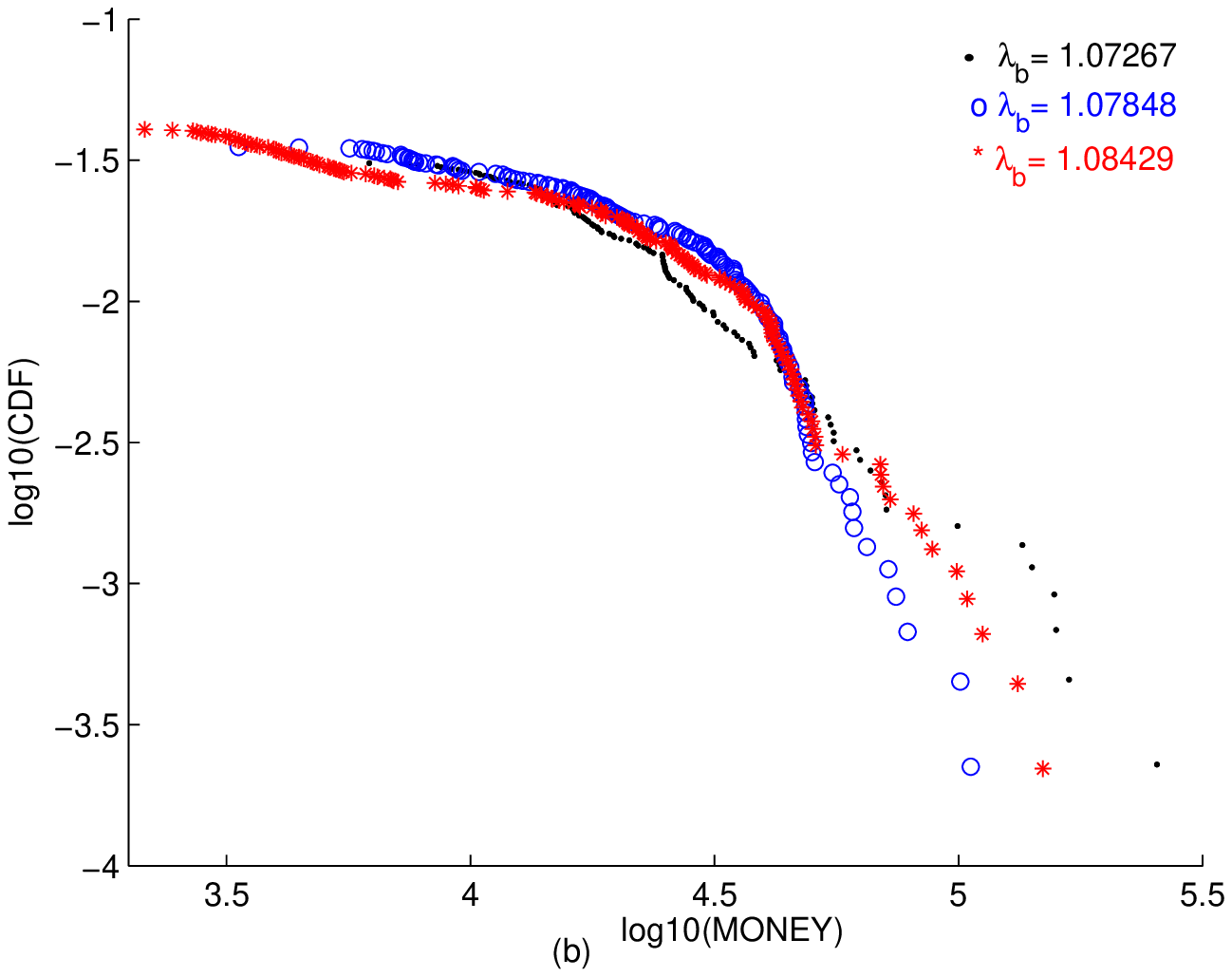}}
  \caption{Representations of the final CDF's obtained for chaotic parameters $\lambda_a = 1.032$
  and different values of $\lambda_b$. In (a) the distribution up to $\$2000$ dollars is depicted for 
  simulation cases 1,2,3,4 and 5. In (b) these CDF's are drawn for simulation cases 6,7 and 8.}
  \label{fig3}
\end{figure}

Fig. \ref{fig3}(a) shows the representation of simulation cases $1$ to $5$ in a natural log plot up to a 
range of $\$2000$. It can be appreciated that as $\lambda_b$ increases, the straight shape obtained for 
the symmetric case bends progressively, the probability of finding an agent in the state of poorness increases. 
It also can be seen that for cases 3 and 4, no agent can be found in a middle range of wealth (from $\$1000$ 
to $\$2000$). This
means that the distribution of money is becoming progressively more unequal.

In Fig. \ref{fig3}(b) the CDF's for simulation cases $6$ to $8$ are depicted from a range of $\$2000$ and 
in double decimal logarithm plot. Here, a minority of agents reach very high fortunes, what explains, how 
other majority of agents becomes to the state of poorness. The data seems to follow a straight line 
arrangement for case $6$, which resembles a Patero distribution. Cases $7$ and $8$ show two straight line 
arrangements which can also be adjusted to two Pareto distributions of different slopes.

To appreciate these results in a deeper detail, one can consider different economic classes of individuals 
according to their final status of wealth. The evolution of the population of individuals can be tracked 
as $\lambda_b$ increases. Let us consider three economic classes:``poor class'' (with final money from $0$ 
to $\$500$, ``middle class'' (from $\$500$ to $\$2000$) and ``rich class'' (with more than $\$2000$).

\begin{table} [h]
\label{table2}
\begin{center}
\begin{tabular}{|c|c|c|c|c|c|c|c|c|}
   \hline
  CASE & $1$ & $2$ & $3$ & $4$ & $5$ & $6$ & $7$ & $8$\\
  \hline
  Total Money\\
  \hline
  POOR & $8.80\%$ & $9.52\%$ & $7.49\%$ & $2.89\%$ & $1.36\%$ & $0.45\%$ & $0.18\%$ & $0.42\%$\\
  \hline
  MIDDLE & $51.58\%$ & $4.37\%$ & $1.71\%$ & $0.13\%$ & $0.0\%$ & $0.0\%$ & $0.0\%$ & $0.0\%$\\
  \hline
  RICH & $39.62\%$ & $86.11\%$ & $90.80\%$ & $96.98\%$ & $98.64\%$ & $99.55\%$ & $99.82\%$ & $99.58\%$\\
  \hline
  Total Population\\
  \hline
  POOR & $39.15\%$ & $93.60\%$ & $96.71\%$ & $98.46\%$ & $97.80\%$ & $96.92\%$ & $96.47\%$ & $95.94\%$\\
  \hline
  MIDDLE & $47.53\%$ & $5.94\%$ & $2.39\%$ & $0.20\%$ & $0.0\%$ & $0.0\%$ & $0.0\%$ & $0.0\%$\\
  \hline
  RICH & $13.32\%$ & $0.46\%$ & $0.90\%$ & $1.35\%$ & $2.20\%$ & $3.08\%$ & $3.53\%$ & $4.06\%$\\
  \hline
\end{tabular}
\end{center}
\caption{Distributions of total traded money and total active agents in different social classes depending 
on the simulation case.}
\end{table}

Table \ref{table2} shows how this society is becoming more unequal as $\lambda_b$ increases. The middle class 
even disappears for $\lambda_b\geq 1.061050$. The rich gets richer as the asymmetry of the chaotic selection 
of agents increases and the final amount of money of this class is almost the total money in the system.

What is happening here is, that the asymmetry of the chaotic map is selecting a set of agents preferably as 
winners for each transaction ($j$ agents). While others, with less chaotic luck become preferably 
looser ($i$ agents).

\begin{figure} [h]
  \centerline{
    \includegraphics[width=0.5\textwidth]{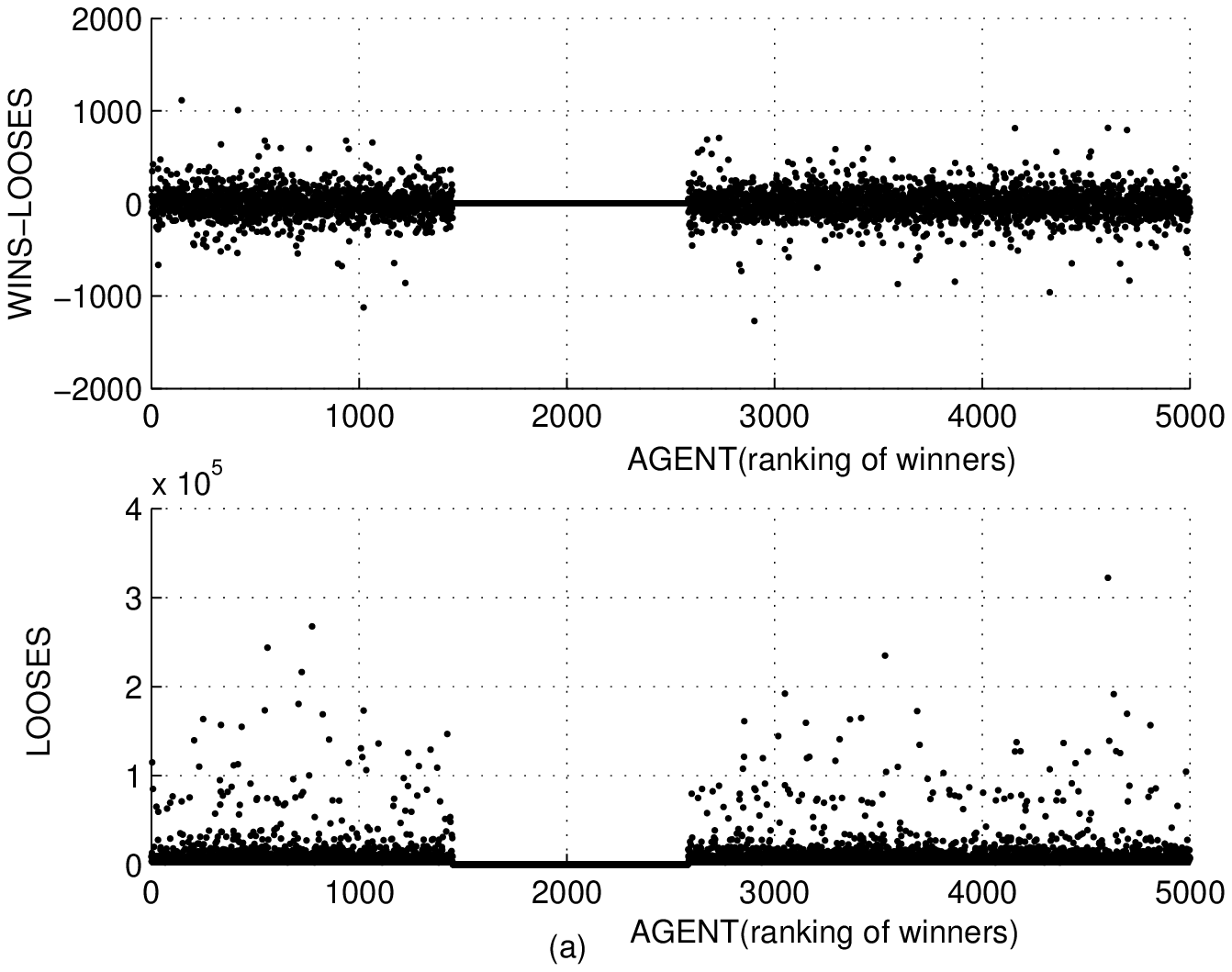} \hskip 3mm 
	\includegraphics[width=0.5\textwidth]{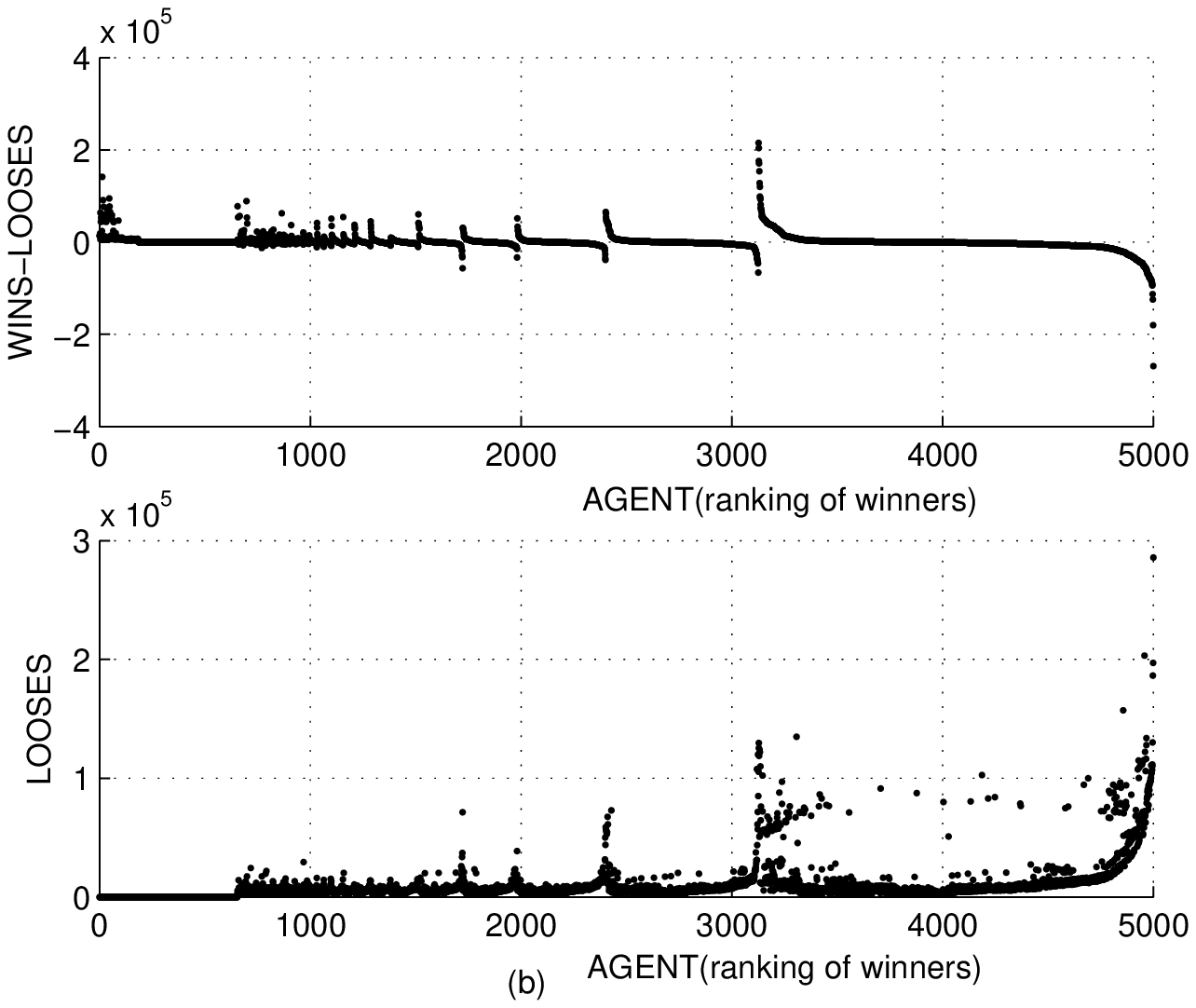}}
  \caption{Representation of the role of all agents after all the interactions. Agents are arranged 
  in descending order according to their final wealth. The upper graphic shows the total number of wins over 
  looses of an agent. The bottom graphic shows the number of times an agent has been selected as $i$ agent or looser. (a) Simulation case 1, $\lambda_b= 1.032$. (b) Simulation case 8, $\lambda_b= 1.084290$.}
  \label{fig4}
\end{figure}

Fig. \ref{fig4} shows, in number of interactions, the times an agent has been a looser (bottom graph) and 
the difference of winning over losing times (top graph). The $x$ axis shows the ranking of agents ordered 
by its final money, in a way so that, agent number $0$ is the richest of the community and agent number $5000$ 
is in the poorest range.

Fig. \ref{fig4}(a) depicts the symmetric case, where $\lambda_a= \lambda_b= 1.032$. Here, the number of 
wins and looses is uniformly distributed among the community. There also is a range of agents that don't 
interact ($1133$ agents), this can be seen clearly in the figures now. In this case, the chaotic selection 
of agents show no particular preference and the final distribution becomes the exponential. Similar to 
traditional simulations with random agents ~\cite{yakovenko2007}.

Fig. \ref{fig4}(b) shows the same magnitudes for case 8, where $\lambda_b= 1.084290$ and the asymmetry 
is maximum. Here it can be seen that there is a group of agents in the range of maximum richness that 
never loose. The chaotic selection is giving them maximum luck and this makes them richer and richer at 
every transaction. These are $184$ rich agents (the $4.06\%$ of Table \ref{table2}). A lower range of agents 
than in Fig. \ref{fig4}(a) are passive and never interact ($470$ agents). There is no middle class here, 
and the rest of the community ($4346$ agents) become in state of poorness with a final wealth inferior 
to $\$500$ and of them, $1874$ agents finish with no money at all.

It is also interesting to see in Fig. \ref{fig4}(b) that in the poor class there are agents that have a 
positive difference of wins over looses, but amazingly they are poor anyway. Consequently, one can deduce 
that they are also bad luck guys. They are $j$ agents in most part of their transactions but unfortunately 
their corresponding trading partners ($i$ agents) are poor too, and they can effectively earn low or no 
money in these interactions.

\section{Conclusions}

This work introduces chaotic selection of agents in economic (ideal) gas-like models in a wide range of
simulation conditions, where the symmetry of the chaotic map is controlled. This mechanism is able of breaking
the pairing symmetry of agents $(i,j)\Leftrightarrow(j,i)$ in trading markets. The distributions of money
obtained this way, exhibit a transition from Gibbs to Pareto distributions, as the pairing symmetry is
progressively broken.

More over, it illustrates how a small group of people can be chaotically destined to be very rich, while the
bulk of the population ends up in state of poverty. This may resemble some realistic conditions,
 showing how some individuals can accumulate big fortunes in trading markets, as a natural consequence of
 the intrinsic asymmetric conditions of real economy.

{\bf Acknowledgements} The authors acknowledge some financial support by Spanish grant
DGICYT-FIS200612781-C02-01.

\bibliographystyle{plain}

\bibliography{bibliography-chaos2009}

\end{document}